\documentclass[fleqn,usenatbib]{mnras}

\usepackage{newtxtext,newtxmath}
\usepackage[T1]{fontenc}
\usepackage{ae,aecompl}
\usepackage{graphicx}	% Including figure files
\usepackage{amsmath}	% Advanced maths commands
\usepackage{amssymb}	% Extra maths symbols
\title[Long term optical and infrared variability]{Long term optical and infrared variability characteristics of Fermi Blazars}

\author[Safna P.Z.]{
P.Z. Safna,$^{1}$\thanks{E-mail: pzsafna@gmail.com}
C.S. Stalin$^{2}$, 
Suvendu Rakshit,$^{3}$ and
Blesson Mathew$^{1}$
\\
$^{1}$Department of Physics and Electronics, CHRIST (Deemed to be University), Hosur Road, Bangalore-560029, India\\
$^{2}$Indian Institute of Astrophysics, Block II, Koramangala, Bangalore-560034, India\\
$^{3}$Finnish Centre for Astronomy with ESO (FINCA), University of Turku, Quantum, Vesilinnantie 5, 20014, Finland\\
}

\date{Accepted XXX. Received YYY; in original form ZZZ}

\pubyear{2019}

\begin{document}

\label{firstpage}
\pagerange{\pageref{firstpage}--\pageref{lastpage}}
\maketitle

\begin{abstract}
We present long term optical and near infrared flux variability analysis of 37 blazars detected in the $\gamma$-ray band by the {\it Fermi Gamma-Ray Space Telescope}. Among them, 30 are flat spectrum radio quasars (FSRQs) and 7 are BL Lac objects (BL Lacs). The photometric data in the optical (BVR) and infrared (JK) bands were from the Small and Moderate Aperture Research Telescope System acquired between 2008$-$2018. From cross-correlation analysis of the  light curves at different wavelengths, we did not find significant time delays between variations at different wavelengths, except for three sources, namely PKS 1144$-$379, PKS B1424$-$418 and 3C 273. For the blazars with both B and J-band data, we found that  in a majority of FSRQs and BL Lacs, the amplitude of variability ($\sigma_m$) in the J-band is larger than B-band consistent with the dominance of the non-thermal jet over the thermal accretion disc component. Considering FSRQs and BL Lacs as a sample, there are indications of $\sigma_m$ to increase gradually towards longer wavelengths in both, however, found to be statistically significant only between
B and J-bands in FSRQs. In the B$-$J v/s J colour magnitude diagram, we noticed complicated spectral variability patterns. Most of the objects showed a redder when brighter (RWB) behaviour. Few objects showed a bluer when brighter (BWB) trend, while in some objects both BWB and RWB behaviours were noticed. These results on flux and colour characteristics indicate that the jet emission of FSRQs and BL Lacs is indistinguishable.
\end{abstract}

\begin{keywords}
galaxies:active -- (galaxies:) BL Lacertae objects: general -- galaxies:nuclei -- galaxies:jets
\end{keywords}

%%%%%%%%%%%%%%%%%%%%%%%%%%%%%%%%%%%%%%%%%%%%%%%%%%

%%%%%%%%%%%%%%%%% BODY OF PAPER %%%%%%%%%%%%%%%%%%
\section{Introduction}

Active Galactic Nuclei (AGN) are actively accreting super massive black holes (SMBH) at the centers of massive galaxies \citep{Antonucci,Urry1995PASP..107..803U}. The radiation from AGN is believed to result from the accretion of matter by the SMBH at the center of its host galaxy \citep{Lynden1969Natur.223..690L,
Rees1984ARA&A..22..471R}. Among AGN are the blazars that show 
large amplitude flux variations \citep{Wagner1995ARA&A..33..163W,
Ulrich1997ARA&A..35..445U} over the entire accessible bands of the electromagnetic spectrum. The emission from blazars is dominated by non-thermal emission from their relativistic jets that are aligned close to the observer. They are broadly classified into two categories namely flat spectrum radio quasars (FSRQs) and BL Lac objects (BL Lacs). This division
is from an observational perspective and it is based on the presence of broad emission lines in their optical
spectra with FSRQs having broad emission lines with equivalent width (EW) > 5 \AA. On a more physical ground, it is now thought that FSRQs have radiatively efficient optically thick and geometrically thin accretion disk, while BL Lacs have radiatively inefficient accretion disk that cannot photoionize the 
line emitting clouds \citep{2001A&A...379L...1G,2009MNRAS.396L.105G,2019arXiv191111777G}. 

The broad band spectral energy distribution (SED) of blazars
show a two hump structure. The low energy hump, extending from radio through X-rays is dominated by the optically thin synchrotron emission process by the energetic 
electrons in the jet \citep{Konigl1981ApJ...243..700K,UrryMushosky1982ApJ...253...38U}. The radio/optical emission that contributes to this component of the SED is found to be highly polarized \citep{1980ARA&A..18..321A}. The origin of the high energy hump that spans the X-ray and $\gamma$-ray region is not well understood. In the leptonic scenario \citep{2007Ap&SS.309...95B}, the high energy hump could be produced  by inverse Compton scattering of seed photons by the same electrons 
in the jet that are responsible for the synchrotron emission (synchrotron self Compton; \citealt{1974ApJ...188..353J}) as well as photons that are exterior to the jet (external Compton;\space \citealt{1993ApJ...416..458D,1994ApJ...421..153S}).  However, the nature of the seed photons that take part in the inverse
Compton process is poorly understood.  Alternatively, in the hadronic scenario, the second hump could also be produced by synchrotron radiation from protons that are co-accelerated along with the jet electrons, interaction of relativistic protons with
external radiation fields as well as  proton induced particle 
cascades \citep{2001APh....15..121M,2003APh....18..593M}.
Clues to a better understanding of the emission processes in blazars can be attained by studying correlated flux variations 
between different spectral bands. 

\begin{table*}
\caption{The sample of sources analysed for variability. The details of the sources are from Ackermann et al. (2015)}
\label{table-1}
\begin{tabular}{llrllccl}
\hline
Name & $\alpha_{2000}$       & $\delta_{2000}$ & $z$ & Type    &  SED Type & MJD of observation & Filters \\
\hline
PMN J0017$-$0512 & 00:17:35.82 & $-$05:12:42    & 0.227   & FSRQ   & LSP &  56442 - 57297      & B,V,R,J,K  \\
4C +01.02        & 01:08:38.76 &    01:34:60    & 2.090   & FSRQ   & LSP & 56552 - 57960      & B,R,J  \\
AO 0235+164      & 02:38:38.93 &    16:36:59    & 0.940   & BL Lac & LSP & 54501 - 57265      & B,V,R,J,K  \\
PKS 0250$-$225   & 02:52:47.95 & $-$22:19:25    & 1.419   & FSRQ   & LSP & 56247 - 56994      & B,V,R,J,K   \\
PKS 0301$-$243   & 03:03:26.50 & $-$24:07:11    & 0.260   & BL Lac & HSP & 56171 - 57093      & B,R,J   \\
PKS 0402$-$362   & 04:03:53.70 & $-$36:05:02    & 1.423   & FSRQ   & LSP & 55838 - 57297      & B,V,R,J,K   \\
PKS 0426$-$380   & 04:28:40.42 & $-$37:56:20    & 1.111   & BL Lac & LSP &  56282 - 57791      & B,V,R,J,K   \\
PKS 0454$-$234   & 04:57:03.20 & $-$23:24:52    & 1.003   & FSRQ   & LSP & 55861 - 57144     & B,V,R,J,K   \\
S3  0458$-$02    & 05:01:12.80 & $-$01:59:14    & 2.088   & FSRQ   & LSP &  56192 - 57342     & B,R,J,K   \\
PKS 0502+049     & 05:05:23.18 &    04:59:43    & 0.954   & FSRQ   & LSP &  56357 - 57129      & B,R,J   \\
PKS 0454$-$46    & 05:07:54.60 & $-$61:04:43    & 0.853   & FSRQ   & LSP & 55488 - 57154      & B,R,J   \\
PKS 0528+134     & 05:30:56.36 &    13:31:55    & 2.070   & FSRQ   & LSP &  54501 - 57095     & B,V,R,J   \\
PKS 0537$-$441   & 05:38:50.30 & $-$44:05:09    & 0.896   & BL Lac & LSP &  55590 - 57143     & B,R,J,K   \\
PKS 0637$-$75    & 06:35:46.50 & $-$75:16:17    & 0.653   & FSRQ   & LSP & 55489 - 57152     & B,R,J   \\
0736+01          & 07:39:18.03 &    01:37:05    & ---     & FSRQ   & ISP  &  56986 - 57179     & B,R,J   \\
PKS 0805$-$077    & 08:08:15.50 & $-$07:51:10    & 1.837   & FSRQ   & LSP &  56697 - 57186      & B,R,J   \\
BZQ J0850$-$1213 & 08:50:09.60 & $-$12:13:35    & 0.566   & FSRQ   & LSP  &  55496 - 57196     & B,R,J   \\
OJ 287           & 08:54:48.87 &    20:06:31    & 0.306   & BL Lac & LSP & 54501 - 57857     & B,V,R,J,K   \\
PKS 1004$-$217   & 10:06:46.40 & $-$21:59:20    & 0.331   & FSRQ   & ISP & 55588 - 57185     & B,V,R,J   \\
1059$-$1134      & 10:59:12.43 & $-$11:34:23    & ---     & BL Lac & LSP & 55941 - 57180     & B,R,J   \\
PKS 1144$-$379   & 11:47:01.40 & $-$38:12:11    & 1.048   & FSRQ   & LSP & 55590 - 57180     & B,V,R,J,K   \\
3C 273           & 12:29:06.69 &    02:03:09    & 0.158   & FSRQ   & LSP & 54501 - 57856     & B,V,R,J,K   \\
PKS 1244$-$255   & 12:46:46.80 & $-$25:47:49    & 0.638   & FSRQ   & LSP & 55599 - 57191     & B,R,J   \\
3C 279           & 12:56:11.10 & $-$05:47:22    & 0.536   & FSRQ   & LSP &  54501 - 57953     & B,V,R,J,K   \\
PKS B1406$-$076  & 14:08:56.48 & $-$07:52:27    & 1.494   & FSRQ   & LSP & 54501 - 57867     & B,V,R,J,K   \\
PKS B1424$-$418  & 14:27:56.30 & $-$42:06:19    & 1.522   & FSRQ   & LSP & 55942 - 57915     & B,V,R,J,K   \\
PKS 1510$-$08    & 15:12:50.53 & $-$09:05:59    & 0.360   & FSRQ   & LSP & 54501 - 57919     & B,V,R,J,K   \\
PKS 1622$-$29    & 16:26:06.02 & $-$29:51:27    & 0.185   & FSRQ   & LSP & 54501 - 57150     & B,V,R,J,K   \\
PKS 1730$-$13    & 17:33:02.70 & $-$13:04:50    & 0.902   & FSRQ   & LSP & 54501 - 57197     & B,V,R,J,K   \\
PKS 1954$-$388   & 19:57:59.80 & $-$38:45:06    & 0.630   & FSRQ   & LSP & 55490 - 57341     & B,R,J   \\
2023$-$07        & 20:25:40.66 & $-$07:35:53    & 1.388   & FSRQ   & LSP & 57270 - 57964     & B,R,J   \\
PKS 2052$-$47    & 20:56:16.30 & $-$47:14:48    & 1.489   & FSRQ   & LSP & 55063 - 57187     & B,V,R,J,K   \\
PKS 2142$-$75    & 21:47:12.70 & $-$75:36:13    & 1.139   & FSRQ   & LSP &  55296 - 57193     & B,V,R,J,K   \\
PKS 2155$-$304   & 21:58:52.07 & $-$30:13:32    & 0.117   & BL Lac & HSP & 54603 - 57964     & B,V,R,J,K   \\
3C 454.3         & 22:53:57.75 &    16:08:54    & 0.859   & FSRQ   & LSP & 54640 - 57964     & B,V,R,J,K   \\
PKS 2326$-$502   & 23:29:20.88 & $-$49:55:41    & 0.518   & FSRQ   & LSP &  56108 - 57200     & B,R,J,K   \\
PMN J2345$-$1555 & 23:45:12.46 & $-$15:55:08    & 0.621   & FSRQ   & LSP & 55873 - 57200     & B,R,J,K   \\
\hline
\end{tabular}
\end{table*}

Blazars have been extensively studied for flux variability in multiple wavebands
\citep{Bonning2012ApJ...756...13B,2017ApJ...835..275R,2017MNRAS.466.3309R,2018MNRAS.480.5517L,2018ApJS..237...30M}, 
yet, we do not have a clear understanding of the cause of flux variations in them. In addition to flux variations, blazars also exhibit spectral variations. Available observational results in literature point to complex spectral variability behaviours in 
them \citep{Bonning2012ApJ...756...13B,2018ApJS..237...30M,2019MNRAS.486.1781R}. Analysis of colour magnitude trends in monitoring data of blazars has revealed complex patterns such as bluer-when-brighter (BWB) trend \citep{2009MNRAS.399.1357S}, 
redder when brighter (RWB) trend \citep{2019ApJ...887..185S}, both BWB and RWB trends \citep{2019MNRAS.486.1781R} and no/weak spectral change with brightness \citep{2003A&A...402..151R}. To 
better understand the complex flux and spectral behaviour of blazars it is important to analyse near simultaneous data on a large number of blazars that spans over many years. With this objective we have carried out a systematic analysis of the photometric monitoring data on a total of 37 blazars to characterize their flux and spectral variability behaviour as well as correlations if any between flux variations between different wavelengths. The sample  and data are described in Section 2. The analysis is discussed in Section 3. The results are discussed in Section 4 followed by the summary in the final section. 

\section{Sample and Data }
\begin{figure*}
\includegraphics[width=0.8\paperwidth]{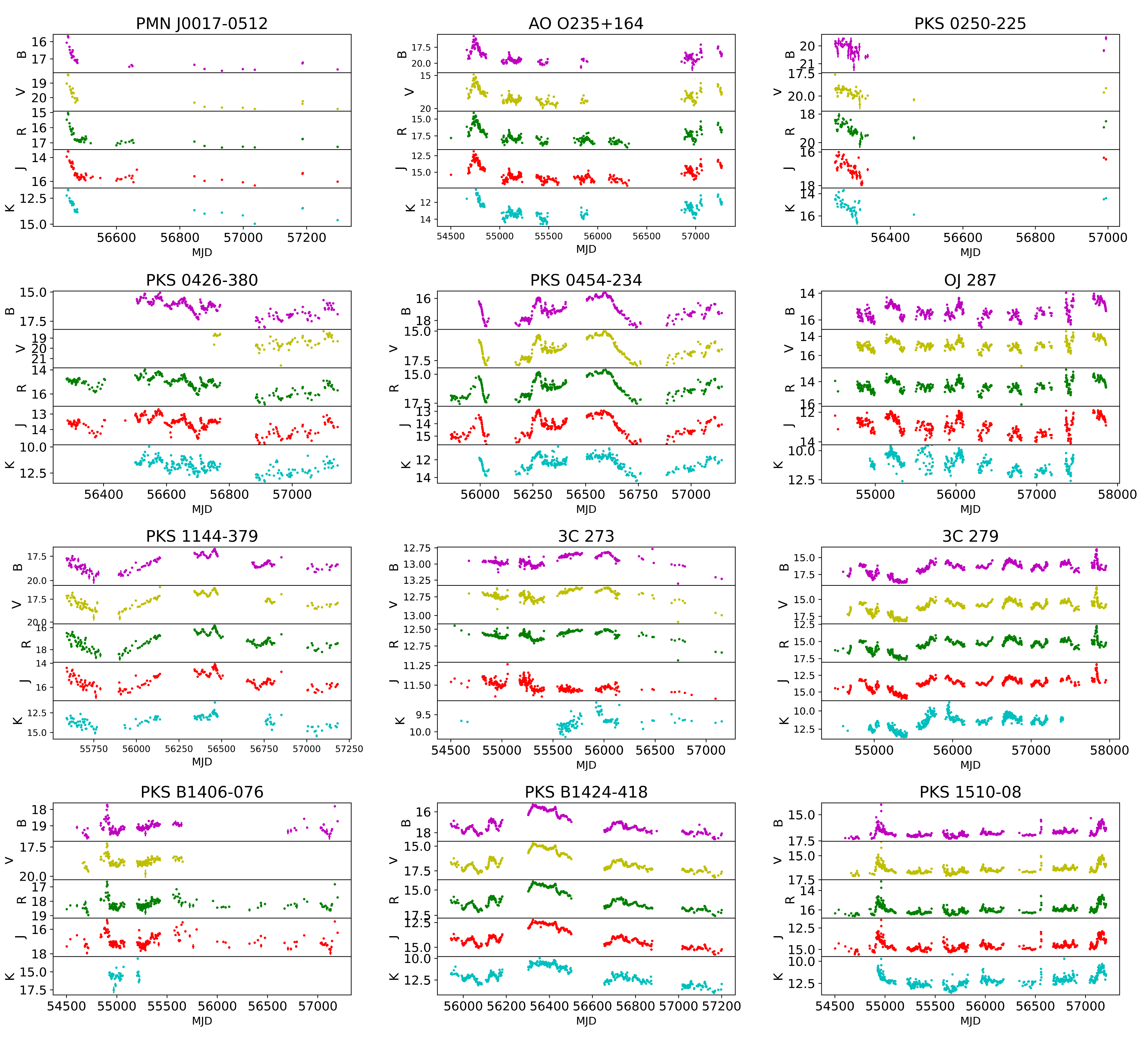}

\caption{Multi-wavelength light curves (magnitude v/s MJD) of the blazars studied in this work. Different colours represent different frequencies. B (Violet), V (Yellow), R (Green) are the optical bands and J (Red) \& K (Blue) are the IR bands. The names of the objects are given on each panel.}
\label{figure-1}
\end{figure*}

\begin{figure*}
\includegraphics[width=0.8\paperwidth]{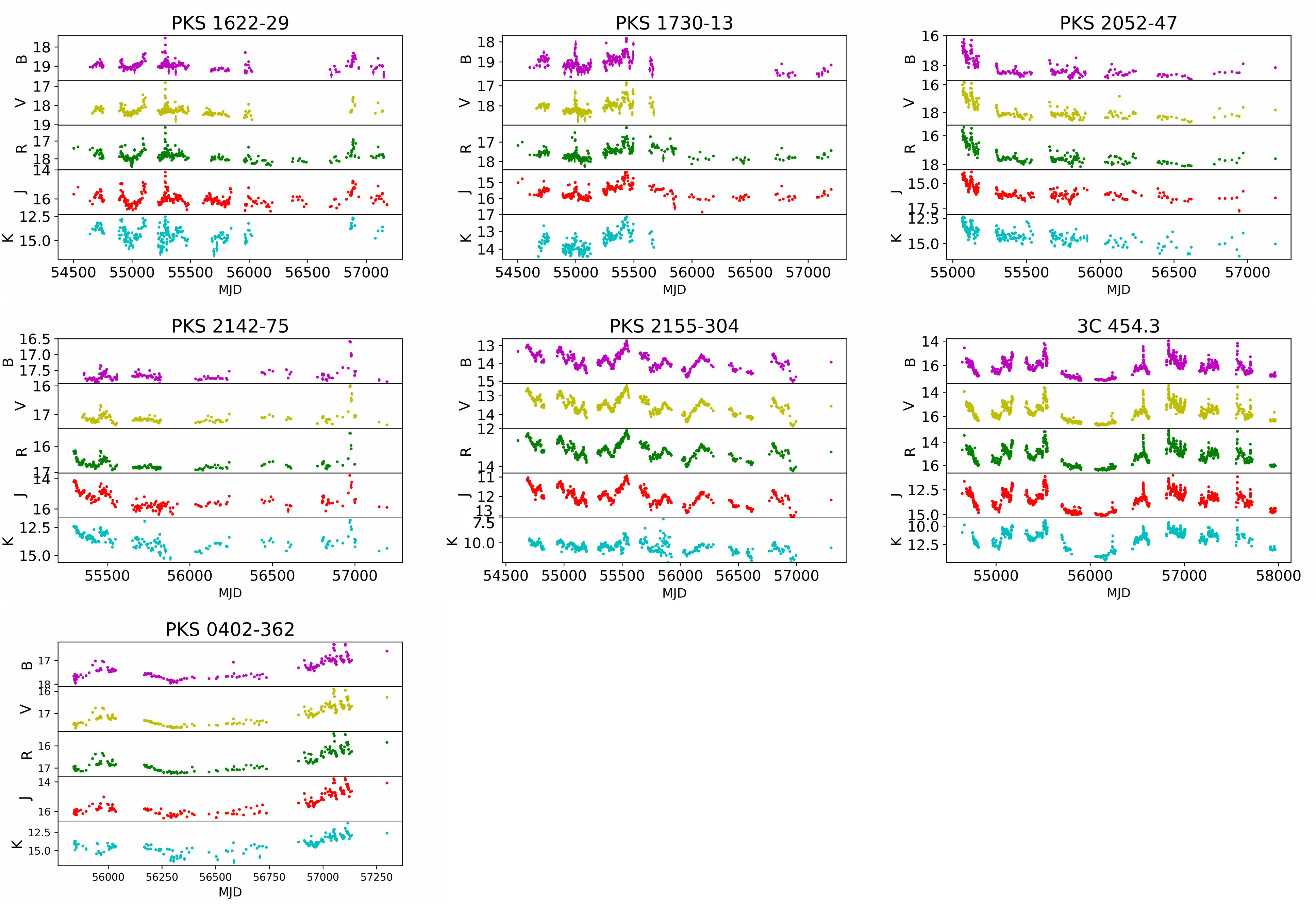}

\caption{Multi-wavelength light curves (magnitude v/s MJD) of the blazars studied in this work. Different colours represent different frequencies. B (Violet), V (Yellow), R (Green) are the optical bands and J (Red) \& K (Blue) are the IR bands. The names of the objects are given on each panel.}
\label{figure-2}
\end{figure*}

\begin{figure*}
\includegraphics[width=0.8\paperwidth]{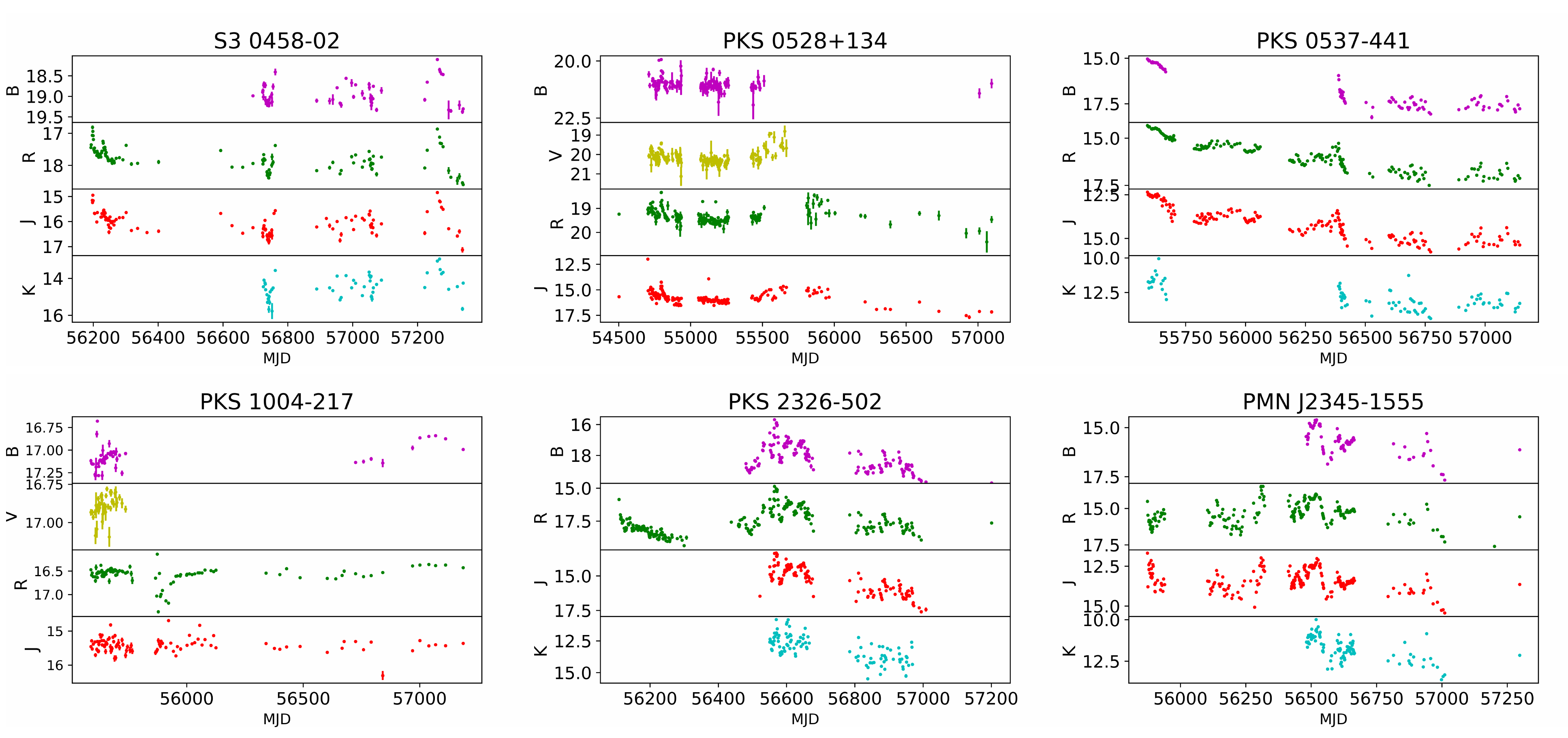}

\caption{Multi-wavelength light curves (magnitude v/s MJD) of the blazars.  Different colours represent different frequencies. B (Violet), V (Yellow), R (Green) are the optical bands and J (Red) \& K (Blue) are the IR bands. The names of the objects are given on each panel.}
\label{figure-3}
\end{figure*}

\begin{figure*}
\includegraphics[width=0.8\paperwidth]{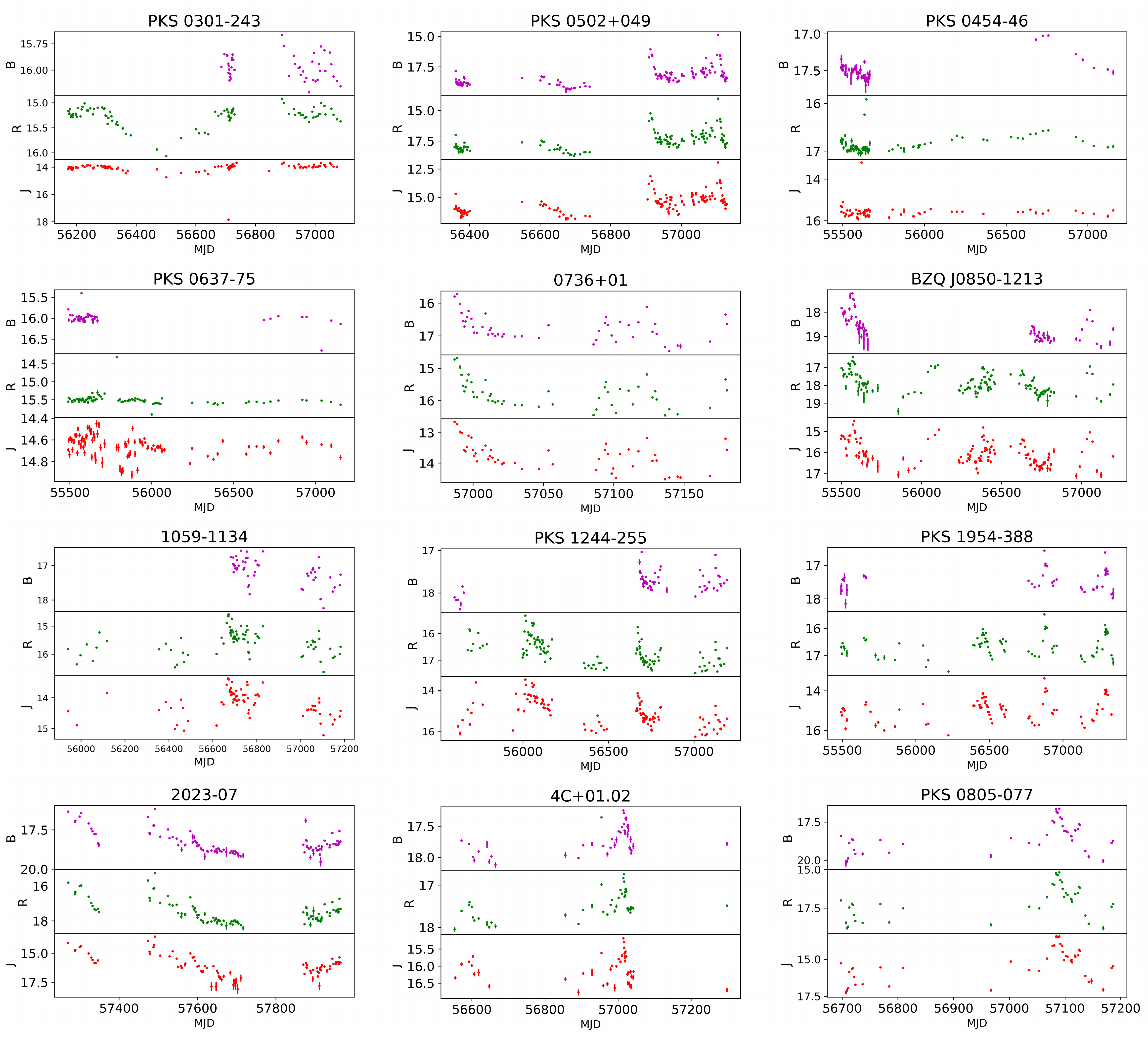}

\caption{Multi-wavelength light curves (magnitude v/s MJD) of the blazars studied in this work. Different colours represent different frequencies. B (Violet), V (Yellow), R (Green) are the optical bands and J (Red) \& K (Blue) are the IR bands. The names of the objects are given on each panel.}
\label{figure-4}
\end{figure*}

Our sample of sources for this study was selected from the list of blazars that 
are routinely monitored by the Small and Moderate Aperture Research 
Telescope System (SMARTS\footnote{http://www.astro.yale.edu/smarts/fermi}, \space\citealt{Bonning2012ApJ...756...13B})
in the optical and near infrared bands. They are part of a list of $\gamma$-ray 
emitting blazars regularly monitored by SMARTS. Firstly, we selected all objects from SMARTS 
and restricted our analysis to only those objects that satisfy the criteria of 
the availability of greater than 40 data points in at least any two optical 
bands and one infrared band during the period 2008 $-$ 2018.  
This constraint leads us to a total  of 37 blazars of which
30 are FSRQs and 7 are BL Lacs. For 19 sources 
in our sample, we have data in all the five photometric bands namely B, V, R, J and K. For 2 sources we have
data in B, V, R and J bands. Data in three bands B, R and J are available for 12 sources in our
sample and the remaining 4 sources have data in four bands, namely B, R, J and K.
The details of the objects that were analysed for flux variations in this work are given in Table \ref{table-1}. 
From visual inspection of each of the light curves, we noticed few photometric points with large error bars. To avoid
such points in the analysis of the light curves, we calculated the median of the photometric errors in each light curve and 
removed those points that have photometric errors larger than ten times the median error.
The final multi-band light curves of the objects are given in 
Figures \ref{figure-1}, \ref{figure-2}, \ref{figure-3} and \ref{figure-4}.

%It is frequently split into subsections, such as Section~\ref{sec:maths} below{}.

\section{Analysis}
\subsection{Flux variability}
%\label{sec:maths} % used for referring to this section from elsewher
To characterise the flux variability nature of the sources, we calculated the intrinsic amplitude of 
variability\space($\sigma_{m}$) as outlined in \cite{Rakshit2017ApJ...842...96R}. The $\sigma_{m}$ was 
calculated from the observed light curves after subtracting the errors in the photometric measurements as
follows (see also \citealt{2007AJ....134.2236S})

\begin{equation} 
   \Sigma =\sqrt{\frac{1}{n-1} \sum_{i=1}^{N}(m_{i}-<m>)^2},
\end{equation}
where <m> is the weighted mean of the magnitude in each band and $\sigma_{m}$ is given by,

\begin{equation}
 \sigma_{m} =
  \begin{cases}
    \sqrt{\Sigma^2 - \epsilon^2} &   \text{if $\Sigma\ge \epsilon$,} \\
         0, &  \text{otherwise}
  \end{cases} 
\end{equation}
$\Sigma^2$ is the variance of the light curve and $\epsilon$ represents the errors in the
photometric measurements and it is calculated directly from individual measurement  errors $\epsilon_{i}$ as,

\begin{equation}
  \epsilon^2 = \frac{1}{N}\sum_{i=1}^{N}\epsilon_{i}^2 
\end{equation}

The values of $\sigma_{m}$ calculated for the sources along with the number of 
data points available in each band are given in Table \ref{table-2}.  Considering 
individual objects, $\sigma_m$ in the infrared J and K bands is larger
than the optical B-band for about 80\% of the objects.  The mean  
$\sigma_m$ values along with their standard deviations in different bands for 
FSRQ and BL Lac as a population are given in Table \ref{table-3}. 
From Table \ref{table-3}, there are indications that the average $\sigma_m$ in 
the infrared J and K bands for the blazars studied here are larger than the 
optical bands for both the FSRQ and BL Lac population, with the trend of increasing $\sigma_m$ 
towards longer wavelengths. However, given the large standard deviations to the 
mean $\sigma_m$ given in Table \ref{table-3}, it is difficult to make 
a strong claim of $\sigma_m$ increasing from shorter towards longer wavelengths
for both FSRQ and BL Lac population. To ascertain the difference in the mean 
$\sigma_m$ between different wavelengths, we did a two sample Kolmogorov-Smironv (KS) test.  
We restricted this test to B, R and J-bands, due to the availability of $\sigma_m$ 
measurements for all the sources in these bands.
For BL Lacs, between B and R-bands, we found a D-statistics of 0.143 and a
null hypothesis (no difference in $\sigma_m$ between B and R bands) probability ($p$) 
of 1.00.  Between B v/s J (R v/s J) we found D-value of 0.286\space (0.286) and null hypothesis
$p$ value of 0.883\space(0.883). Thus, for BL Lacs as a population, the observed hint of increased
variability towards longer wavelengths is not statistically significant.
Similarly, for FSRQs, between B and R-bands, our KS test gave a D-statistic of 0.167 and
$p$ of 0.760.  For variations between B and J-bands, we found a D-value of 0.400 and a 
$p$ of 0.011. And, between R and J-bands, the obtained D and $p$ values are 0.233 and 0.342 respectively. Thus, the two sample KS test conducted between the distributions
of $\sigma_m$ between B and J-bands in FSRQs confirms that the two distributions are indeed different, and therefore, the difference in mean $\sigma_m$ between J and B-band is
statistically significant. Also, the differences in the mean $\sigma_m$ between (B,R) and (R,J) are statistically insignificant.  Similarly in all the bands (except the K-band) the mean $\sigma_m$ in BL Lacs is larger than that of FSRQs (see Table \ref{table-3}). Here too, as the standard deviations are larger, we did a KS test. For B-band, we found a D-statistics and p-value (null hypothesis probability of no difference in the $\sigma_m$ between BL Lacs and FSRQs) of 0.324 and 0.504 respectively. Similarly, for V-band (R-band) we found D and p values of 0.412 (0.224) and 0.519 (0.901) respectively. For J-band we found a D and p of 0.314 and 0.543, while for K-band the D and p values are 0.289 and 0.834 respectively. Thus, from KS test we found that the distribution of $\sigma_m$ between FSRQs and BL Lacs in all the bands is indistinguishable.  Analysis similar to the one carried out in this work is needed on a larger number of FSRQs and BL Lacs to confirm if BL Lacs indeed show $\sigma_m$ larger than FSRQs on month to year like time scales.

% \begin{tiny}
% \usepackage{graphicx}
\begin{table*}
\caption{Results of the analysis of variability. The numbers in parenthesis are the number of data points corresponding to each band.}
\label{table-2}
\begin{tabular}{lcccccll}
\hline
NAME & B & V & R & J & K & Spectral variability & Type \\ 
\hline
PMN J0017$-$0512   & 0.498 (54)   & 0.617 (45)  & 0.528 (81)  & 0.675 (85)  & 0.699 (51)  & RWB       &  FSRQ \\
4C+01.02         & 0.233 (42)   &  ---        & 0.306 (44)  & 0.403 (42)  & ---         & RWB       & FSRQ \\
AO 0235+164      & 0.962 (221)  & 1.007 (239) & 1.003 (301) & 1.084 (333) & 0.858 (238) & RWB,BWB   & BL Lac \\
PKS 0250$-$225   & 0.346 (49)   & 0.500 (55)  & 0.547 (55)  & 0.442 (56)  & 0.664 (48)  & RWB       & FSRQ \\
PKS 0301$-$243   & 0.114 (46)   & ---         & 0.179 (97)  & 0.432 (93)  & ---         & RWB       & BL Lac \\
PKS 0402$-$362   & 0.370 (175)  & 0.412 (174) & 0.425 (169) & 0.666 (167) & 1.036 (156) & RWB       & FSRQ \\
PKS 0426$-$380   & 0.579 (372)  & 0.564 (73)  & 0.538 (480) & 0.451 (482) & 0.566 (384) & BWB       & BL Lac \\
PKS 0454$-$234   & 0.788 (340)  & 0.761 (339) & 0.765 (381) & 0.703 (369) & 0.714 (336) & BWB       & FSRQ \\
S3 0458$-$02     & 0.283 (57)   & ---         & 0.357 (110) & 0.427 (103) & 0.626 (52)  & RWB        & FSRQ \\ 
PKS 0502+049     & 0.717 (131)  & ---         & 0.778 (131) & 0.844 (135) & ---         & RWB        & FSRQ \\
PKS 0454$-$46    & 0.148 (42)   & ---         & 0.182 (65)  & 0.345 (57)  & ---         & RWB        & FSRQ \\
PKS 0528+134     & 0.281 (117)  & 0.337 (146) & 0.300 (186) & 0.605 (207) & ---         & RWB        & FSRQ \\
PKS 0537$-$441   & 0.967 (91)   & ---         & 0.791 (192) & 0.851 (179) & 0.851 (86)  & RWB        & BL Lac \\
PKS 0637$-$75    & 0.159 (45)   & ---         & 0.151 (85)  & 0.096 (88)  & ---         & RWB        & FSRQ \\
0736+01          & 0.377 (52)   & ---         & 0.412 (52)  & 0.452 (53)  & ---         & RWB        & FSRQ \\
PKS 0805$-$077      & 0.954 (43)   & ---         & 1.042 (42)  & 1.080 (44)  & ---         & RWB        & FSRQ \\
BZQ J0850$-$1213 & 0.522 (76)   & ---         & 0.571 (144) & 0.564 (132) & ---         & RWB        & FSRQ \\
OJ 287           & 0.522 (573)  & 0.507 (576) & 0.476 (578) & 0.535 (525) & 0.665 (417) & RWB        & BL Lac \\
PKS 1004$-$217   & 0.121 (53)   & 0.070 (42)  & 0.173 (98)  & 0.228 (93)  & ---         & RWB        & FSRQ \\
1059$-$1134      & 0.366 (52)   & ---         & 0.430 (83)  & 0.417 (79)  & ---         & No trend   & BL Lac \\
PKS 1144$-$379   & 0.692 (173)  & 0.743 (148) & 0.662 (184) & 0.591 (181) & 0.705 (132) & No trend   & FSRQ \\
3C 273           & 0.074 (316)  & 0.061 (318) & 0.056 (325) & 0.062 (306) & 0.183 (121) & RWB        & FSRQ \\
PKS 1244$-$255   & 0.261 (69)   & ---         & 0.501 (190) & 0.677 (167) & ---         & RWB        & FSRQ \\
3C 279  & 0.958 (778)  & 0.944 (771) & 0.905 (780) & 0.886 (702) & 0.894 (584) & RWB,BWB  & FSRQ \\
PKS B1406$-$076  & 0.320 (203)  & 0.359 (179) & 0.348 (221) & 0.503 (213) & 0.643 (47)  & RWB        & FSRQ \\
PKS B1424$-$418  & 0.881 (428)  & 0.903 (425) & 0.893 (426) & 0.891 (423) & 0.911 (412) & RWB        & FSRQ \\
PKS 1510$-$08    & 0.410 (598)  & 0.469 (596) & 0.485 (600) & 0.599 (588) & 0.650 (529) & RWB        & FSRQ \\
PKS 1622$-$29    & 0.235 (257)  & 0.232 (251) & 0.296 (276) & 0.423 (313) & 0.755 (234) & RWB        & FSRQ \\
PKS 1730$-$13    & 0.345 (243)  & 0.297 (226) & 0.329 (282) & 0.390 (280) & 0.502 (211) & RWB        & FSRQ \\
PKS 1954$-$388   & 0.320 (47)   &  ---        & 0.397 (92)  & 0.588 (94)  & ---         & RWB        & FSRQ \\
2023$-$07        & 0.687 (103)  &  ---        & 0.633 (101) & 0.940 (85)  & ---         & RWB        & FSRQ \\
PKS 2052$-$47    & 0.600 (208)  & 0.636 (208) & 0.606 (215) & 0.705 (223) & 0.731 (229) & RWB        & FSRQ \\
PKS 2142$-$75    & 0.183 (161)  & 0.198 (165) & 0.215 (196) & 0.532 (205) & 0.644 (194) & RWB        & FSRQ \\
PKS 2155$-$304   & 0.410 (462)  & 0.400 (460) & 0.396 (462) & 0.409 (407) & 0.507 (398) & No trend   & BL Lac \\
3C 454.3         & 0.621 (883)  & 0.651 (849) & 0.712 (868) & 0.849 (856) & 1.020 (689) & RWB        & FSRQ \\
PKS 2326$-$502   & 0.845 (169)  &   ---       & 0.898 (279) & 0.950 (139) & 0.961 (107) & RWB    & FSRQ \\
PMN J2345$-$1555 & 0.646 (91)   &   ---       & 0.727 (217) & 0.664 (214) & 0.708 (99)  & RWB        & FSRQ \\
\hline
\end{tabular}
\end{table*}

\begin{table*}
\caption{Mean variability characteristics of the blazars in optical and infrared bands}
\label{table-3}
\begin{tabular}{llllll}
\hline
class  & B & V & R & J & K \\ \hline
BL Lac & 0.56 $\pm$ 0.29 & 0.62 $\pm$ 0.23 & 0.54 $\pm$ 0.25 & 0.60 $\pm$ 0.25 & 0.69 $\pm$ 0.14 \\
FSRQ   & 0.46 $\pm$ 0.26 & 0.48 $\pm$ 0.26 & 0.51 $\pm$ 0.25 & 0.59 $\pm$ 0.24 & 0.72 $\pm$ 0.19 \\ \hline
\end{tabular}
\end{table*}

\begin{table*}
\caption{Results of the linear least squares fit to the colour magnitude diagram.}
\label{table-4}
\begin{tabular}{lrrrl}
\hline
NAME & Slope & Intercept & R & p  \\ 
\hline
PMN J0017$-$0512   &  $-$0.32$\pm$0.03 &  6.75$\pm$0.41 & $-$0.89  & 4.33 $\times$ 10$^{-19}$ \\
4C+01.02         &  $-$0.68$\pm$0.07 & 12.56$\pm$1.12 & $-$0.90  & 1.31 $\times$ 10$^{-14}$ \\
AO 0235+164      &     0.18$\pm$0.03 & 1.75$\pm$0.37  &    0.52  & 1.11 $\times$ 10$^{-06}$ \\
                 &  $-$0.43$\pm$0.04 & 10.73$\pm$0.70  & $-$0.69  & 4.43 $\times$ 10$^{-20}$ \\
PKS 0250$-$225   &  $-$0.87$\pm$0.18 & 17.91$\pm$3.11 & $-$0.61  & 4.98 $\times$ 10$^{-06}$  \\
PKS 0301$-$243   &  $-$0.92$\pm$0.23 & 14.86$\pm$3.21 & $-$0.48  & 1.05 $\times$ 10$^{-3}$  \\
PKS 0402$-$362   &  $-$0.48$\pm$0.01 &  9.24$\pm$0.21 & $-$0.92  & 4.48 $\times$ 10$^{-69}$  \\
PKS 0426$-$380   &     0.19$\pm$0.02 & 0.07$\pm$0.28  & 0.43     & 7.65 $\times$ 10$^{-18}$ \\
PKS 0454$-$234   &     0.09$\pm$0.02 & 1.55$\pm$0.24  & 0.26     & 3.11 $\times$ 10$^{-06}$ \\
S3 0458$-$02     &  $-$0.49$\pm$0.04 & 10.69$\pm$0.66 & $-$0.85  & 5.14 $\times$ 10$^{-15}$ \\
PKS 0502+049     &  $-$0.22$\pm$0.03&  6.18$\pm$0.50 & $-$0.79 & 2.07 $\times$ 10$^{-28}$  \\ 
PKS 0454$-$46    &  $-$1.23$\pm$0.36 & 21.12$\pm$5.68 & $-$0.66 & 1.03 $\times$ 10$^{-05}$ \\
PKS 0528+134     &  $-$0.88$\pm$0.08 & 19.19$\pm$1.33 & $-$0.81 & 1.59 $\times$ 10$^{-25}$  \\
PKS 0537$-$441   &  $-$0.11$\pm$0.01 &  4.05$\pm$0.17 & $-$0.69 & 1.76 $\times$ 10$^{-13}$ \\
PKS 0637$-$75    &  $-$1.19$\pm$0.23 & 18.80$\pm$3.42 & $-$0.75 & 4.33 $\times$ 10$^{-08}$ \\
0736+01          &  $-$0.22$\pm$0.04 & 6.06$\pm$0.56  & $-$0.68  & 7.45 $\times$ 10$^{-08}$ \\
PKS 0805$-$077    &  $-$0.13$\pm$0.02 & 5.13$\pm$0.35  & $-$0.65 & 2.20 $\times$ 10$^{-06}$ \\
BZQ J0850$-$1213 &  0.05$\pm$0.14 & 1.99$\pm$2.20  &    0.30  & 0.24 \\
                 &  $-$0.57$\pm$0.11 & 11.94$\pm$1.78 & $-$0.67  & 5.57 $\times$ 10$^{-07}$ \\
OJ 287           &  $-$0.20$\pm$0.03 & 5.08$\pm$0.36  & $-$0.24 & 5.17 $\times$ 10$^{-08}$ \\
PKS 1004$-$217   &  $-$1.06$\pm$0.09 & 17.96$\pm$1.45 & $-$0.81 &  1.01 $\times$ 10$^{-12}$ \\
1059$-$1134      &  $-$0.05$\pm$0.05 & 3.62$\pm$0.76  & $-$0.16 & 0.26  \\
PKS 1144$-$379   &     0.07$\pm$0.04 & 1.69$\pm$0.60  &    0.12 & 0.14 \\
3C 273           &  $-$1.92$\pm$0.11 & 23.54$\pm$1.25 & $-$0.77  & 1.34 $\times$ 10$^{-56}$  \\
PKS 1244$-$255   &  $-$0.52$\pm$0.05 & 10.37$\pm$0.77 & $-$0.84 & 3.15 $\times$ 10$^{-16}$ \\
3C 279           & 0.10$\pm$0.02 & 1.49$\pm$0.24 &    0.31 & 3.55 $\times$ 10$^{-14}$  \\
                 &  $-$0.48$\pm$0.07 & 10.40$\pm$1.12 & $-$0.49 & 1.49 $\times$ 10$^{-06}$ \\
PKS B1406$-$076  &  $-$0.51$\pm$0.05 & 10.87$\pm$0.85 & $-$0.79 & 1.48 $\times$ 10$^{-40}$ \\
PKS B1424$-$418  &  $-$0.03$\pm$0.01 & 3.54$\pm$0.12 & $-$0.15 & 2.54 $\times$ 10$^{-03}$ \\
PKS 1510$-$08    &  $-$0.37$\pm$0.02 & 7.51$\pm$0.24 & $-$0.86 & 2.64 $\times$ 10$^{-167}$  \\
PKS 1622$-$29    &  $-$0.63$\pm$0.04 & 13.11$\pm$0.64 & $-$0.91 & 8.20 $\times$ 10$^{-92}$ \\
PKS 1730$-$13    &  $-$0.50$\pm$0.06 &  11.31$\pm$0.87 & $-$0.56 & 5.51 $\times$ 10$^{-19}$ \\
PKS 1954$-$388   &  $-$0.59$\pm$0.04 & 11.29$\pm$0.58 & $-$0.93 & 2.33 $\times$ 10$^{-18}$ \\
2023$-$07        &  $-$0.41$\pm$0.06 &  8.79$\pm$0.97 & $-$0.62 & 2.74 $\times$ 10$^{-10}$ \\
PKS 2052$-$47    &  $-$0.07$\pm$0.06 &  3.43$\pm$0.90 & $-$0.28 & 6.00 $\times$ 10$^{-2}$ \\
                 &  $-$0.66$\pm$0.07 & 12.93$\pm$1.21 & $-$0.68 & 6.49 $\times$ 10$^{-19}$ \\
PKS 2142$-$75    &  $-$0.86$\pm$0.05 & 12.52$\pm$0.72 & $-$0.97 & 3.87 $\times$ 10$^{-86}$ \\
PKS 2155$-$304   &  $-$0.01$\pm$0.01 &  1.95$\pm$0.11 & $-$0.00 & 0.98 \\
3C 454.3         &  $-$0.30$\pm$0.01 &  6.72$\pm$0.13 & $-$0.85 & 2.63 $\times$ 10$^{-233}$ \\
PKS 2326$-$502   &     0.18$\pm$0.14 &  $-$0.13$\pm$1.98 &    0.09 & 0.53 \\
                 &  $-$0.30$\pm$0.02 &  7.48$\pm$0.40 & $-$0.75 & 5.91 $\times$ 10$^{-17}$ \\
PMN J2345$-$1555 &  $-$0.06$\pm$0.02 &  3.08$\pm$0.21 & $-$0.26 & 1.49 $\times$ 10$^{-2}$ \\
\hline
\end{tabular}
\end{table*}

\subsection{Colour Variability}
\begin{figure*}
 \includegraphics[width=0.8\paperwidth]{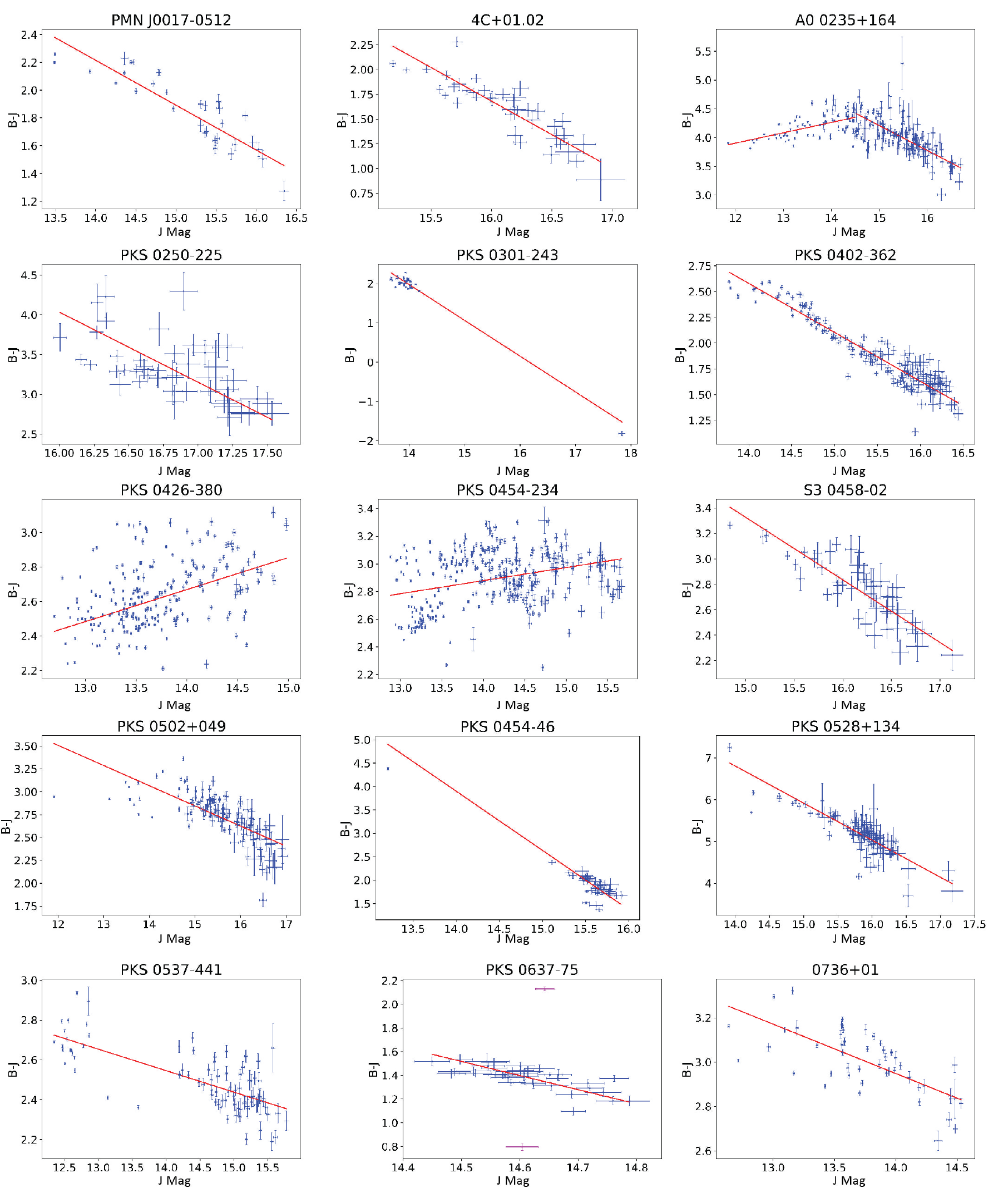}

\caption{Correlation between (B$-$J) color and J-band brightness of the blazars. The solid line is the linear
\label{figure-5}
least squares fit to the data. The names of the objects are given on each panels.}
\end{figure*}

\begin{figure*}
\includegraphics[width=0.8\paperwidth]{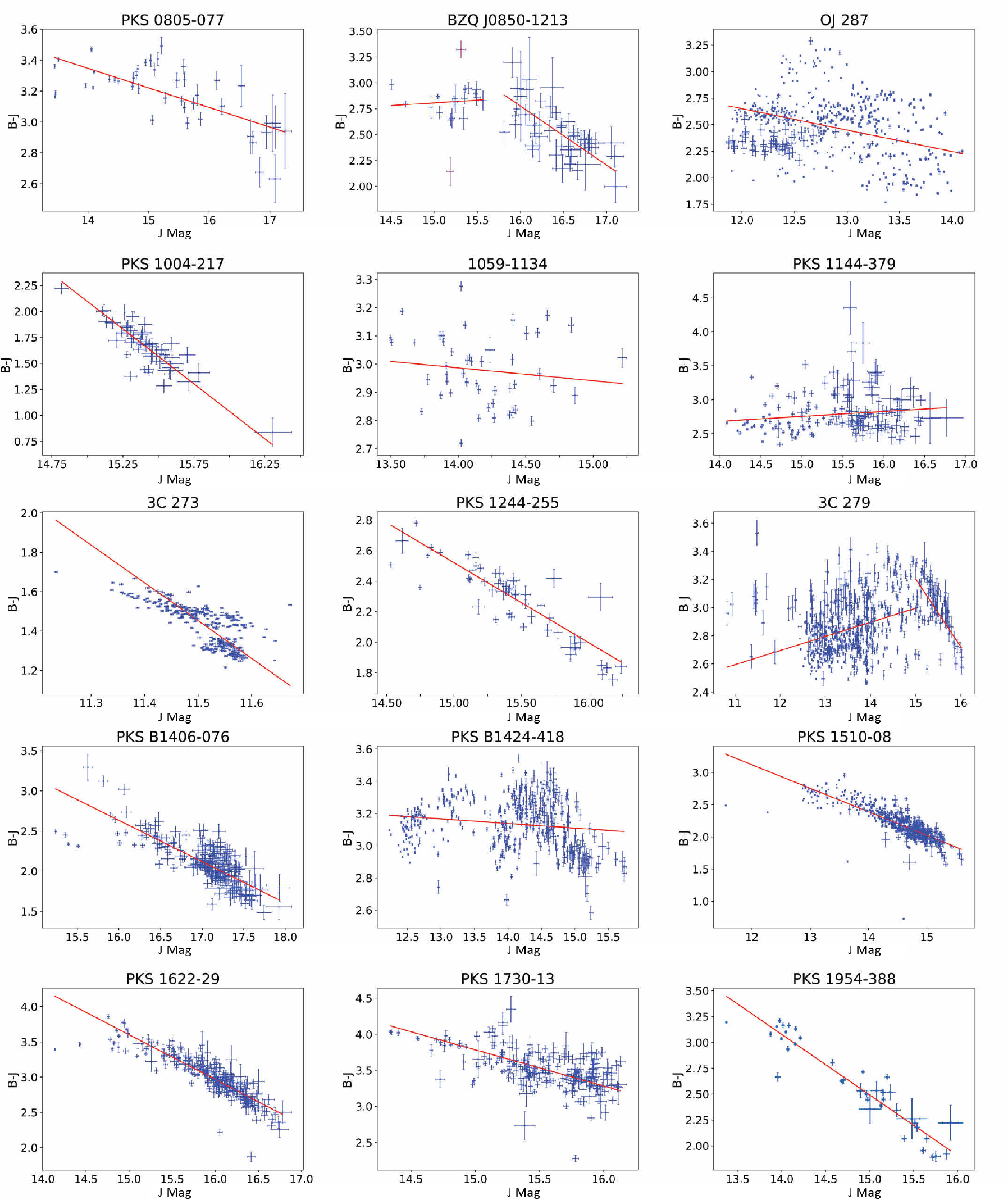}
 
\caption{Correlation between (B$-$J) color and J-band brightness of the blazars. The solid line is the linear
least squares fit to the data. The names of the objects are given on each panels.}
\label{figure-6}
\end{figure*}

\begin{figure*}
\includegraphics[width=0.8\paperwidth]{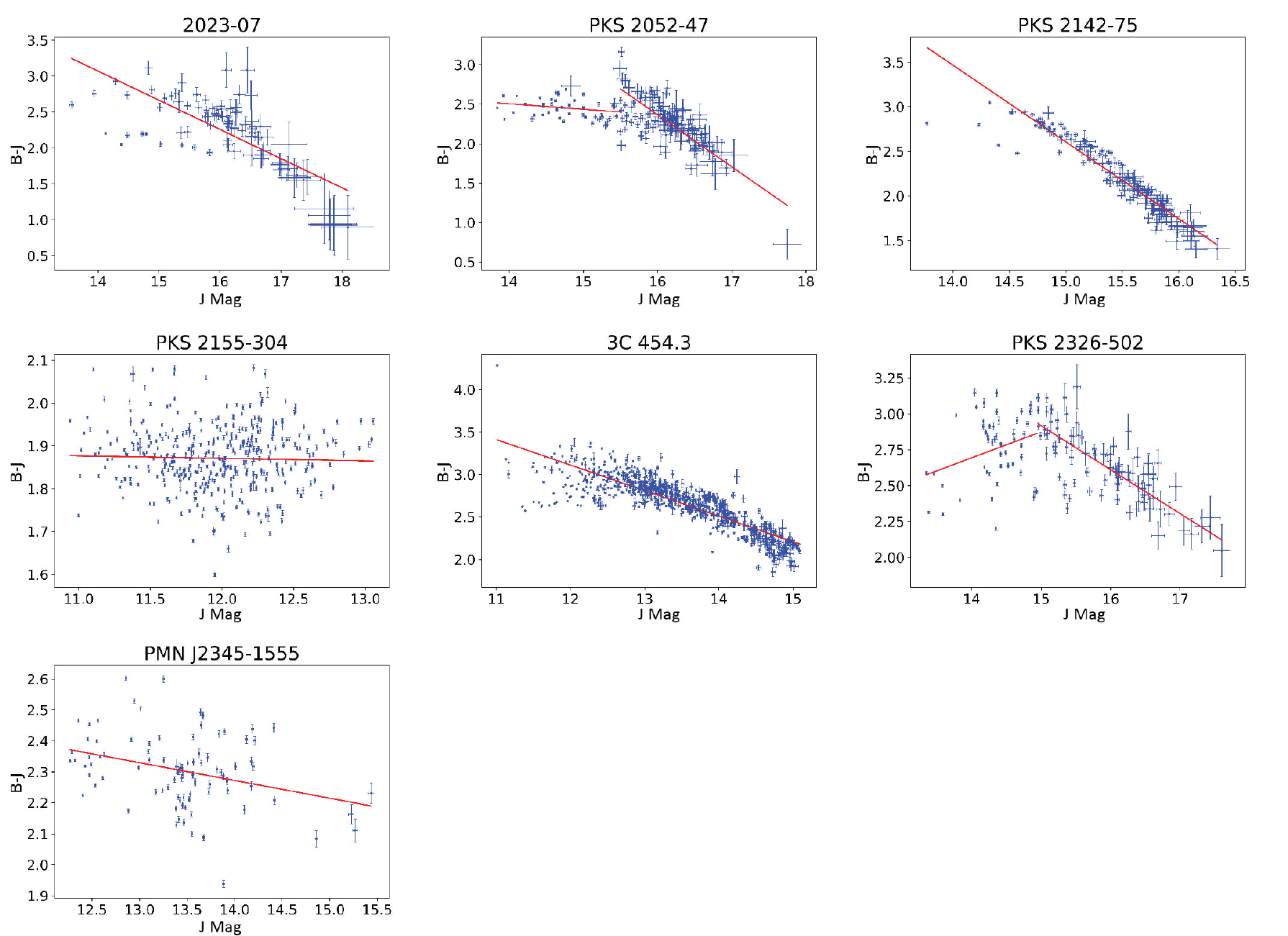} 
\caption{Correlation between (B$-$J) color and J-band brightness of the blazars. The solid line is the linear
least squares fit to the data. The names of the objects are given on each panels.}
\label{figure-7}
\end{figure*}

In addition to flux variations, blazars are also known to show spectral variations. The photometric measurements
in BVRJK bands analysed here have contributions from both thermal emission coming from the accretion disk  and non-thermal emission coming from the relativistic jet. Both these processes contribute differently to the emission in different bands. Thus by studying spectral variations, one can distinguish the different components contributing to the observed flux variations. To characterise the nature of spectral variations in our sample of sources, we generated colour (B$-$J) -  magnitude (J-band) diagrams. Linear
least squares fit was carried out on the data points in the colour-magnitude diagram by properly taking into account the errors in both colour and magnitude. For this we used BCES method \footnote{https://github.com/rsnemmen/BCES} \citep{1996ApJ...470..706A,2012Sci...338.1445N} of orthogonal least squares to find the slope and intercept. A positive slope in the colour-magnitude diagram is an indication of a BWB 
trend, while a negative slope is an indication of a RWB trend. The generated colour-magnitude diagrams for all the objects in our sample, along with the linear least squares fit are given in Figures
\ref{figure-5}, \ref{figure-6} and \ref{figure-7}. The results of the linear least squares fit to the 
colour magnitude diagram are given in Table \ref{table-4}. 
We considered BWB color-magnitude variation in 
a source if the Spearman rank correlation coefficient is positive and the probability $p$ of no correlation is lesser than 0.05. Similarly, a source is considered to have shown a RWB trend if values of the Spearman rank correlation coefficient is negative and $p$ is lesser than 0.05. Using the above criteria we found that a large fraction of the sources in our sample (~80\%) showed a RWB trend. No spectral change with brightness was noticed in three sources, while another two sources showed a BWB trend. In two sources namely A0 0235+164 and 3C 379, we noticed both BWB and RWB trend. They showed a RWB trend upto a certain J-band brightness, and for magnitudes brighter than that they showed a BWB behaviour. For two sources PKS 0637$-$75 and BZQ J0850$-$1213 two deviant point each (shown as purple points in \ref{figure-5} and \ref{figure-6} respectively) were excluded from the fit, as inclusion of those points gave large errors in the derived slope and intercept. These observations indicate that complex spectral variations
with brightness are invariably present in FSRQs and BL Lacs.

\subsection{Multi-wavelength cross-correlation}   
All the sources analysed in this work are highly variable in both optical and infrared bands. 
To check for correlation between the flux variations in different bands we used 
the discrete correlation function (DCF; \citealt{Edilson1988ApJ...333..646E,whiteprtrson1994PASP..106..879W}) and the interpolated cross-correlation function (ICCF) 
techniques \citep{1986ApJ...305..175G,1987ApJS...65....1G}.
The DCF method calculates the correlation function from unevenly sampled data. For two light 
curves, say, $x_i$ and $y_j$ this method calculates a set of unbinned discrete correlation functions (UDCFs) given as
\begin{equation}
UDCF_{ij} = \frac{(x_i - \overline{x}) (y_j - \overline{y})}  {\sqrt{(\sigma_x^2 - e_x^2)(\sigma_y^2 - e_y^2)}}
\end{equation}
here $\overline{x}$ and $\overline{y}$ are the mean of the two light curves, $\sigma_x$ and $\sigma_y$ are
the standard deviations of the two light curves and $e_x$ and $e_y$ are the errors in each data point. For 
any time lag ($\tau$), DCF  is taken as the average of all UDCFs having the same $\tau$. Taking the average of 
M pairs of UDCFs for which $\tau - \Delta\tau/2 \le \Delta\tau_{ij} < \tau + \Delta\tau/2$, we get
\begin{equation}
DCF(\tau) = \frac{1}{M} \sum UDCF_{ij}
\end{equation}
The error in each DCF bin is given as
\begin{equation}
\sigma(\tau) = \frac{1}{M-1} {\sum[UDCF_{ij} - DCF(\tau)]^2}^{1/2}
\end{equation}

To estimate the lag we used 
the centroid ($\tau_{cen}$) of the cross-correlation function (CCF) defined as

\begin{equation}
\tau_{cen} = \frac{\sum_i \tau_i CCF_i} {\sum_i CCF_i}
\end{equation}

\begin{figure*}
\includegraphics[width=0.8\paperwidth]{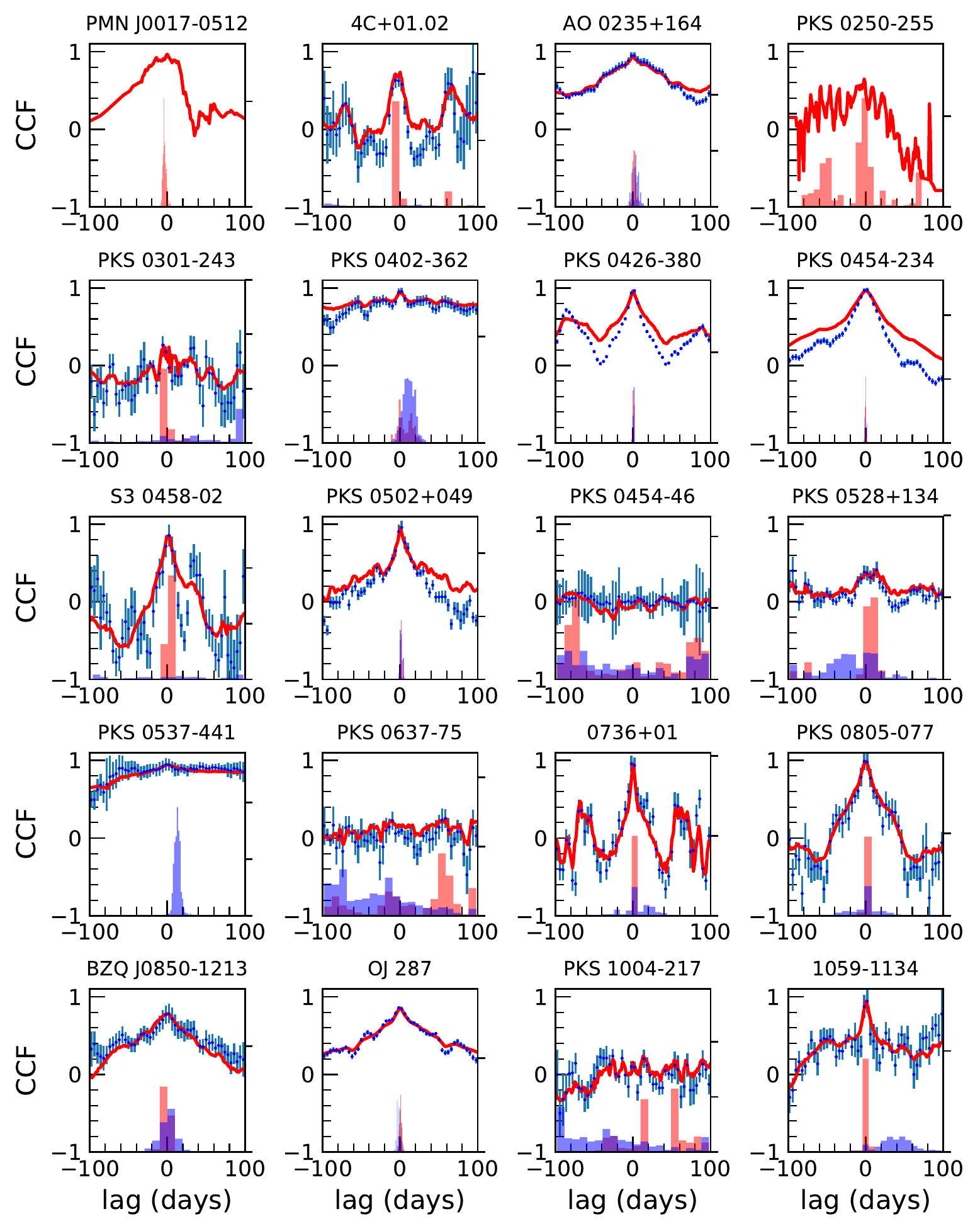}
\caption{Correlation between the optical B-band and infrared J-band flux variations. The solid lines refer to 
ICCF while the blue filled circles with error bars refer to DCF. The distribution of the centroids of the cross-correlation
functions are given in blue (ICCF) and orange (DCF) respectively.}
\label{figure-8}
\end{figure*}

\begin{figure*}
\includegraphics[width=0.8\paperwidth]{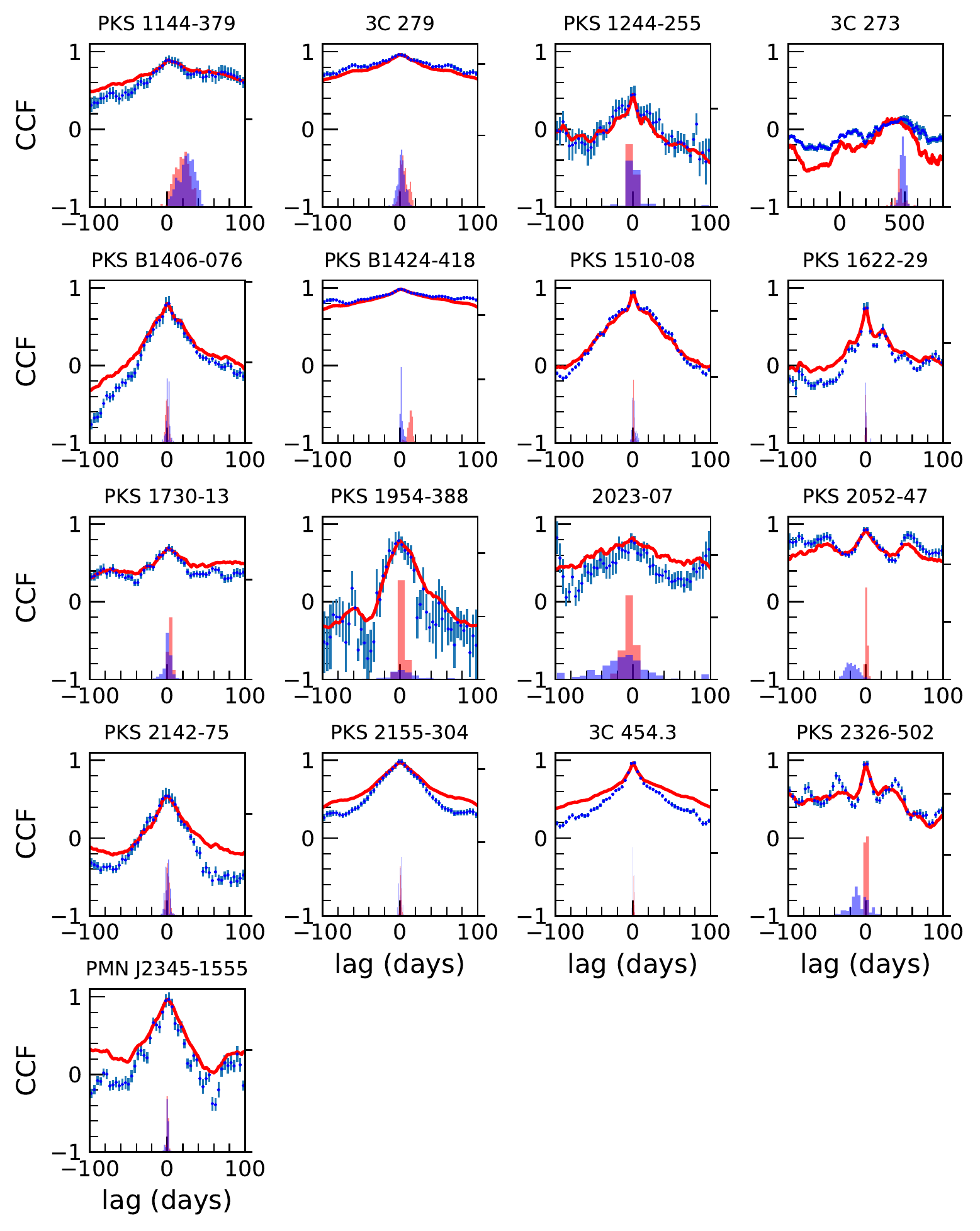}

\caption{Correlation between the optical B-band and infrared J-band flux variations. 
The symbols and the distributions have the same meaning as in Fig. \ref{figure-8}}
\label{figure-9}
\end{figure*}

 For calculating the centroid, we included those points in the CCF that are within 80\% of the CCF peak. To get the errors in the CCF, we followed the Monte Carlo approach based on flux randomization (FR) and random subset selection (RSS) outlined in \cite{petersonccfA1998PASP..110..660P} that takes into account the time sampling of the light curves and the randomness in the flux measurements. For each pair of light curves using both FR and RSS, we carried out 10,000 iterations, and for each iteration we determined the $\tau_{cen}$. Using these values, we generated the distribution of CCF values. The median of the CCF distribution is taken as the value of the lag. As the distribution of $\tau_{cen}$ obtained from 10,000 iterations is non-Gaussian, the upper ($\Delta \tau_{up}$) and lower ($\Delta \tau_{low}$)  uncertainties in $\tau_{cen}$ were determined such that 15.87\% of iterations have $\tau > \tau_{cen} + \Delta \tau_{up}$ and 15.87\% of iterations have $\tau < \tau_{cen} - \Delta \tau_{low}$. The upper and lower error thus obtained in $\tau_{cen}$  corresponds to $\pm$ 1 $\sigma$ for a function that has a Gaussian distribution. The above procedures to obtain the lag and the associated error using DCF method was also repeated  using ICCF technique, with the exception that the light  curves were first brought to uniform time binning using linear interpolation. Thus, for all the sources we have two measurements of lag, one using the DCF method and the other one using ICCF method. The cross-correlation plots obtained between B and J bands is shown in Figures \ref{figure-8} and \ref{figure-9}. For majority of the sources, we found no lag between the flux variations in B and J-bands. Only for three sources, namely, PKS 1144$-$379 (lag = {${19.3^{+8.9}_{-10.7}}$} days), PKS B1424$-$418 (lag= ${13.4^{+2.0}_{-3.4}}$ days) and 3C 273 (lag = ${461.0^{+37.6}_{-30.2}}$ days), we found statistically significant lags between flux variations in B and J-bands.

\section{Discussion}
\subsection{Flux variability}
The long term optical/UV and infrared flux variations observed in blazars could be due to intrinsic processes, extrinsic processes or a combination of both. Internal
physical process include  those related to accretion disk caused by change in accretion rate \citep{2008MNRAS.387L..41L} and those related to the relativistic jet such as blobs of emission moving through the helical magnetic fields of the jet \citep{2008Natur.452..966M}, the propagation of 
shock waves in helical jets \citep{2013ApJ...768...40L} or changes in the direction and speed of the jet flow \space \citep{2016ApJ...820...12P}. Alternatively, the observed long term flux changes in blazars could also be caused by extrinsic geometrical effects associated with a change in
our viewing angle to the moving blob of emission, that leads to variable Doppler boosting of the emitted radiation \citep{1992A&A...259..109G,1992A&A...255...59C,2009A&A...501..455V,2010A&A...510A..93L}. 
The observed brightness in the UV, optical and the infrared bands in blazars is a combination of blue thermal emission from the accretion disk and red relativistically beamed non-thermal emission from their relativistic jets
\citep{2008bves.confE...9P}. The 
emission from the disk peaks in the UV band (the big blue bump at around 
3000 \AA) and its contribution decreases towards optical and infrared 
wavelengths. In the infrared band, the thermal emission is thought to come
from the dusty molecular torus that reprocesses the photons from the broad
line region and the accretion disk into the infrared region \citep{1993ARA&A..31..473A,2008bves.confE...9P}. As the thermal emission comes from a larger region in the accretion disk, they are expected to vary less 
than the variations caused from the jet. The blazars studied here are $\gamma$-ray loud sources detected by the {\it Fermi Gamma-ray Space Telescope}
\citep{2015ApJ...810...14A}, and are believed to have strong relativistic jets whose non-thermal emission 
dominates over the thermal accretion disk emission most of the time. Therefore, in this scenario, it is quite natural to expect $\sigma_m$ to increase with wavelength. This
is what is observed on individual sources in this work, where in majority of the objects (about 80\%), $\sigma_m$ in the infrared J and K bands is larger than that in the 
optical B-band irrespective of FSRQs or BL Lacs. 
Considering FSRQs and BL Lacs as a population,  the mean $\sigma_m$ is found to progressively increase towards longer wavelengths in both FSRQs and BL Lacs 
(see Table \ref{table-3}) though the standard deviations on each of the mean $\sigma_m$ values are very large. A two sample KS test was therefore carried out on the distribution of $\sigma_m$ values between different sets of filters such as (B,R), (R,J) and (B,J)
for both FSRQs and BL Lacs. Statistically significant difference was only found between the 
distribution of $\sigma_m$ values in B and J-bands for FSRQs. 
%Also, from
%Table \ref{table-3} it is likely that BL Lacs show larger $\sigma_4$ than FSRQs, however
%again they are statistically insignificant. 

Among blazars, a prominent accretion disk 
component is mostly seen in FSRQs that have broad emission lines and a radiatively efficient accretion disk relative to BL Lacs that have weak emission lines and radiatively inefficient accretion disk 
\citep{2001A&A...379L...1G,2009MNRAS.396L.105G,2019arXiv191111777G}. However, it is not always true that accretion disk emission is prominent in FSRQs, 
as there are instances wherein at increased activity states, the jet emission dominates the accretion disk emission making the accretion disk emission 
non-distinguishable in the broad band SED. 
Similarly, in BL Lacs too there are instances where the strong disc component is prominent in 
their broad band SED \citep{2008A&A...480..339R} particularly during their faint activity states. 
Irrespective of the conspicuous presence of accretion disk component in the broad band
SEDs of both FSRQs and BL Lacs, the observed optical and infrared emission in most of our sample
of $\gamma$-ray loud blazars studied here is unambiguously dominated by the red jet emission.
Therefore, the increased $\sigma_m$ towards longer wavelengths seen in majority of the 
blazars studied here is understood by the dominance of the jet emission in them.
Dividing the 37 blazars studied here based on the position of the synchrotron peak in their 
broad band SED, we have 33 low synchrotron peaked sources (LSPs), 2 intermediate synchrotron peaked sources (ISPs) and 2 high synchrotron peaked sources (HSPs). In both the ISPs and HSPs too, the amplitude of variability in the infrared J-band is found to be larger than the optical B-band. Considering the average variability properties of the sample of FSRQs and BL Lacs as a whole, we found hints of increased variability amplitude towards longer 
wavelengths in both the populations which again supports the dominance of jet emission in both the population of blazars.  
The number of BL Lacs used in this study is only seven, therefore, similar analysis on a large number of FSRQs and
BL Lacs are needed  to statistically confirm the indications observed from the mean $\sigma_m$ values shown in Table \ref{table-3}. Considering, individual sources, 
in a small fraction of the objects in the sample (that includes FSRQs and BL Lacs) the amplitude of variability is found to increase towards shorter wavelengths (see Table \ref{table-2}) which could be due to the low jet activity of the sources and the emission from the accretion
disk becoming prominent over the jet emission. 

Based on an analysis of a small sample of blazars utilising data from SMARTS, but spanning only two years \cite{Bonning2012ApJ...756...13B} found FSRQs to show increased variability towards longer wavelengths, while they found BL Lacs not to show larger variability amplitude towards longer wavelengths. Similarly, \cite{2015RAA....15.1784Z} on analysis of a large number of FSRQs and BL Lacs found FSRQs to show larger $\sigma_m$ in 
the infrared J-band relative to the optical R-band. In the case of BL Lacs they found the $\sigma_m$ to be larger in the R-band relative to J-band. However, in this work, from an analysis of about 10 years of data, in a larger number of FSRQs and BL Lacs we found hints of  increased $\sigma_m$ towards longer wavelengths in both FSRQs and BL Lac, though statistically significant difference
was only found between B and J-bands in FSRQs.  
Our results for FSRQs are in general in agreement with that of \cite{Bonning2012ApJ...756...13B} and \cite{2015RAA....15.1784Z}, however our
 results on BL Lacs are at odds with what is available in 
literature \citep{Bonning2012ApJ...756...13B,2015RAA....15.1784Z}.

Considering the sample of FSRQs and BL Lacs studied in this work as a whole, we found BL Lacs to have  larger mean $\sigma_m$ in all the bands analysed here except K-band, though the standard deviations in the mean $\sigma_m$ values are larger. Given the large standard deviation, it is difficult to conclusively say
on strong statistical ground that BL Lacs show larger $\sigma_m$ than FSRQs. Given this caveat, it is likely that both FSRQs and BL Lacs have the same intrinsic jet variability. However, as BL Lacs have small inclination than FSRQs, the difference in the $\sigma_m$ values of optical and infrared variability  between FSRQs and BL Lacs with BL Lacs showing more $\sigma_m$ compared
to FSRQs could be due to BL Lacs viewed at smaller angles than FSRQs and subsequently subjected to higher
Doppler boosting. Therefore, the difference in the mean $\sigma_m$ between FSRQs and BL Lacs found here
is in line with what is expected in the Unification scheme of radio-loud AGN, but to place
the results that are found to be indicative here on a firm footing, this analysis needs
to be extended on a large sample of FSRQs and BL Lacs.

\subsection{Colour Variability}
The observed optical, infrared emission in blazars is the sum of the blue thermal emission from the accretion disk and red non-thermal emission from the relativistic jet. The variations seen in the optical and infrared emission from blazars are often accompanied by colour\space(spectral) variations. Studies of the correlation between colour and brightness variations in blazars can give clues to the relation between accretion disk and jet emission in blazars.
However the correlation between colour and brightness in blazars is highly debated in literature. Reports available  in literature indicate that among blazars, in BL Lacs the colour becomes bluer with increasing brightness (BWB;\space\citealt{2002PASA...19..111D,2003ApJ...590..123V,2004A&A...419...25F,2018ApJS..237...30M,2019MNRAS.484.5633G}), 
while in FSRQs the colour becomes redder with increasing brightness (RWB; \citealt{Bonning2012ApJ...756...13B,2019ApJ...887..185S}). Such a distinct colour magnitude relation between FSRQs and BL Lacs lead to speculate that the jets of FSRQs and BL Lacs are fundamentally different.

However, in addition to the distinct colour magnitude relation between FSRQs and BL Lacs, varied colour magnitude relations too were seen in blazars. For example in the FSRQ
3C 454.3, BWB trend was observed at certain epochs, while RWB trend was observed at certain other epochs and at one epoch both BWB and RWB trends were observed \citep{2019MNRAS.486.1781R}. Also, in another FSRQ, 3C 345, BWB and RWB trends were simultaneously observed 
\citep{2011MNRAS.418.1640W}. In the BL Lac object  S5 0716+714, \cite{2003A&A...402..151R} 
found  the colour to be weakly correlated with brightness with one clear BWB trend at a certain period, while \cite{2009MNRAS.399.1357S} found BWB trend on both 
inter-night and intra-night timescales. A RWB trend arises if the relativistic jet emission dominates over the less 
variable thermal component from the accretion disk \citep{2019ApJ...887..185S}.
Alternatively, a BWB trend is attributed to increased amplitude of variability at shorter wavelengths
\citep{2009MNRAS.399.1357S}. Such a scenario, in the one zone synchrotron emission model could happen due to the injection of fresh electrons that have an energy distribution harder than that of the earlier cooler elections, that causes an increase in flux with
a BWB trend \citep{1998A&A...333..452K,2002PASA...19..138M}. Another explanation for a BWB behaviour is attributed to a change in the Doppler factor \citep{2004A&A...421..103V}. 
According to \cite{2007A&A...470..857P} the BWB colour variation could be due to changes in the 
Doppler factor, that could happen due to geometrical effects on the variation in the viewing angle of a curved and inhomogeneous jet.

Majority of the sources in our sample showed a RWB trend on analysis of their complete data set. In all such sources, the amplitude of variability in the infrared J-band
is larger than the optical B-band except three sources namely PKS 0537$-$441, PKS 0637$-$75 and 3C 273.
Of these three, PKS 0537$-$441 is a BL Lac object while the other two are FSRQs. In two sources in our sample, namely A0 0235+164 and 3C 279 we found  a RWB trend upto a certain J-band brightness beyond which the spectrum
changes shape with increasing brightness. Based on the position of the synchrotron peak, of the 37 blazars, 33 are LSP sources, 2 are HSPs and 2 are ISPs. The RWB trend is invariably found among all 
the blazar sub-classes as well as among BL Lacs and FSRQs.
All the objects studied here are $\gamma$-ray emitters and thus have conspicuous 
jets in them. In this work we found similar flux and colour variability patterns in both FSRQs and BL Lacs. 
This lead us to conclude that the jets in both FSRQs and BL Lacs have a similar 
flux and colour variability behaviour.
\section{Summary}

We carried out an analysis of the multiwavelength (BVRJK) flux variability 
characteristics of a sample of 37 blazars (that 
includes 30 FSRQs and 7 BL Lacs)  taken
from the SMARTS archives using data that spans over 10 years from 2008$-$2018. The main findings are summarized below.

\begin{enumerate}
\item All the sources analysed here showed long term flux variations in both the optical and infrared bands. In a majority of the sources ($\sim$80\%), the amplitude of variations in the infrared J-band is larger than the amplitude of variations in the optical B-band. Considering FSRQs and BL Lacs as a population, there 
are hints that the  amplitude of flux variations in the longer
wavelengths is larger than that at the shorter wavelengths. This behaviour is also
evident among the different sub-classes of blazars, namely LSP, ISP and HSP sources. 
For FSRQs we found statistically significant difference between the flux variations in 
B and J bands. Such increased variations at the longer wavelengths is due to the dominance of 
the jet emission in blazars.

\item Between FSRQs and BL Lacs as a population, the data analysed in this work is 
suggestive (though statistically insignificant) of BL Lacs showing increased amplitude of flux variations in the optical 
and infrared bands than FSRQs. This is understood as the jets of BL Lacs are 
viewed at much smaller angles than FSRQs and thereby
subjected to large variability amplitude due to Doppler boosting.

\item We found close correlations between colour and brightness changes with 
a majority of FSRQs and BL Lacs predominantly showing a RWB trend.

\item The observed flux and colour variations in this work, lead to conclude 
that the relativistic jets in FSRQs and BL Lacs are similar.
\end{enumerate}

\section{Acknowledgements}
We thank the referee for his/her critical comments that helped to improve the manuscript. This paper has made use of up-to-date SMARTS optical/near-infrared light curves that are available at www.astro.yale.edu/smarts/glast/home.php. Safna, P.Z., thanks CHRIST (Deemed to be University), for providing the facilities to carry out this work. The help by Arun Roy, Ujjwal Krishnan, Sabik PS, Raghu Krishnan and Shakhil PG is also thankfully acknowledged. 

\section{Data availability}
The photometric data underlying this article are publicly available from the Small and Moderate
Aperture Research Telescope System (SMARTS\footnote{http://www.astro.yale.edu/smarts/fermi}) portal. 

%colleagues, acknowledge funding agencies, telescopes and facilities used etc.

%%%%%%%%%%%%%%%%%%%%%%%%%%%%%%%%%%%%%%%%%%%%%%%%%%


%\bibliographystyle{mnras}


%%%%%%%%%%%%%%%%%%%%%%%%%%%%%%%%%%%%%%%%%%%%%%%%%%

%%%%%%%%%%%%%%%%% APPENDICES %%%%%%%%%%%%%%%%%%%%%

%\appendix



%%%%%%%%%%%%%%%%%%%%%%%%%%%%%%%%%%%%%%%%%%%%%%%%%%

% Don't change these lines
\bsp	% typesetting comment

\label{lastpage}

\end{document}